# Multi-measures fusion based on multi-objective genetic programming for full-reference image quality assessment

Naima Merzougui

*Kasdi Merbah University of Ouargla,
BP 511 Ouargla 30000, Algérie
LESIA laboratory, Mohamed Khider University of
Biskra, PO Box 145 RP 07000 Biskra, Algeria*
merzougui.naima@gmail.com

Leila Djerou

*LESIA laboratory, Mohamed Khider University of
Biskra, PO Box 145 RP 07000 Biskra, Algeria*
djerou.leila@gmail.com

**Abstract:**

*In this paper, we exploit the flexibility of multi-objective fitness functions, and the efficiency of the model structure selection ability of a standard genetic programming (GP) with the parameter estimation power of classical regression via multi-gene genetic programming (MGGP), to propose a new fusion technique for image quality assessment (IQA) that is called Multi-measures Fusion based on Multi-Objective Genetic Programming (MFMOGP). This technique can automatically select the most significant suitable measures, from 16 full-reference IQA measures, used in aggregation and finds weights in a weighted sum of their outputs while simultaneously optimizing for both accuracy and complexity. The obtained well-performing fusion of IQA measures are evaluated on four largest publicly available image databases and compared against state-of-the-art full-reference IQA approaches. Results of comparison reveal that the proposed approach outperforms other state-of-the-art recently developed fusion approaches.*

## I. Introduction

Images have an important role in the media, but their use depends on their quality. They are affected by a wide variety of deformations during acquisition, processing, compression, storage, transmission and reproduction.
There are several methods of evaluating image quality where they are classified into two categories: subjective and objective. The subjective methods of image quality assessment (IQA)
are performed by human subjects, these methods are precise in estimating the visual quality of an image but they take considerable time and requiring a large number of observers, in addition they cannot be automated. On the other hand the objective image quality assessment methods are computer based methods that can automatically predict the perceived image quality. Therefore the objective image quality assessment methods attempt to evaluate the quality of the image in the same way as how humans perceive image quality, they would be relatively quicker and cheaper than subjective assessment.

Depending on the nature of the information required to perform the assessment, quality metrics are classified into three categories [1], [2]:
- Quality metrics with full reference denoted FR (Full Reference) [3], which compare the version of the degraded image with a reference version.
- Reduced reference quality metrics, RR (Reduced Reference) [4], which compare a description of the image to be evaluated with a description of a reference version.
- Non-reference metrics, NR (No Reference) [5], which characterize distortions of the degraded image to be evaluated only from the latter.

In response to the importance of the full-reference metric, a large number of various image and video quality assessment approaches have been proposed.
Some of public image databases, in the area of image quality assessment, are used to compare IQA approaches. They contain reference images, their corresponding distorted images, and ground-truth information obtained from human observers. Information on the perceived quality is reported as mean opinion scores (MOS values) or differential mean opinion scores (DMOS values).
However, quality assessment of images subjected to various types of distortions is still one of the challenging problems in computer vision and image analysis. For this reason, recently, many studies have adopted a new strategy involving different fusion techniques for quantifying image quality. For example, in the most apparent distortion algorithm (MAD) [6] local luminance and contrast masking evaluate high-quality images. In [7], a combination of local and global distortion measures was considered using saliency maps, gradient and contrast information. In [8, 9], scores of MSSIM, VIF and R-SVD were non-linearly combined. Canonical correlation analysis is used to combined SNR, SSIM, VIF, and VSNR in [10], the regularized regression was used to combine up to seven IQA models in [11]. Peng and Li in [12] presented an approach based on conditional Bayesian mixture of expert's model. In that paper, a support vector machines classifier was used for prediction of the type of distortion, and then SSIM, VSNR, and VIF with k-nearest-neighbour regression were fused. An adaptive combination of IQA approaches with an edge-quality based on preservation of edge direction was introduced in [13].Liu and Lin in [33] introduces a fusion measure using a support vector regression approach. In [14], a fusion measure which combines six IQA measures using a neural network was proposed. In [15], kernel ridge regression was used to combine found perceptually meaningful structures and local distortion measurements. A preliminary work with non-linear combination of several IQA measures selected by a genetic algorithm was shown in [16]. In [17], the fusion was formulated as an optimization problem that was solved using the genetic algorithm, which was also responsible for selection of appropriate IQA measures.

In this work, a new fusion technique is formulated as a multi-objective optimization problem which is solved using the Multi-objective genetic programming (MOGP). The proposed technique can automatically select the most significant IQA measures for well-performing fusion, formulate the model structure, and solve the unknown parameters of the regression equation, while simultaneously optimizing for both accuracy and complexity.

The paper is organized as follows; in section 2, we present multi-objective genetic programming (MOGP). In section 3 we describe our new metric MFMOGP, and we discuss the study's evaluation of experimental results. A conclusion of the work is then given is section 4.

## II. Multi-objective genetic programming

Multi-objective genetic programming (MOGP) [18] is an extension of Genetic programming (GP) used for modeling complex nonlinear engineering systems. Genetic programming GP [19] is a programming method inspired by Darwin's theory of evolution. It aims to find, by successive approximations, programs (functions or expressions) best suited to a given task. This is done by using randomly generating a population of computer programs and then breeding together the best performing programs to create a new population (offspring). This process is iterated until the population contains programs that solve the task. Excellent, free to download introduction and review of the GP literature is provided by [20].

In recent years, several advancements for GP have been suggested. Multi-gene genetic programming (MGGP) [21] is of the most recent advancements of that combines the ability of the standard GP in constructing the model structure with the capability of traditional regression in parameter estimation [21]. MGGP is more accurate and efficient than the standard GP for modeling nonlinear problems.

In traditional GP the population is a set of individuals, each of them denoted by a tree structure that is composed of the terminal and function set. The elements of the terminal set $X$ can be input process variables and random constants. The function set codifies a mathematical equation, which describes the relationship between the output $Y$ and a set of input variables $X$.

Based on these ideas, MGGP generalizes GP as it denotes an individual as a structure of trees, also called genes (multigenes), that similarly receives $X$ and tries to predict $Y$.

In MGGP, each prediction of the output variable Y is formed by a weighted output of each of the trees/genes in the multi-gene individual plus a bias term. The mathematical form of the multigene representation can be written as:

$$Y = \sum_{i=1}^{N} d_i G_i(X) + d_0 \quad (1)$$

Where $G_i$ is the value of the $i^{th}$ gene ( it is a function of one or more of the input variables of the terminal set $X$, $d_i$ is the $i^{th}$ weighting coefficient, $N$ is the number of genes, $d_0$ is a bias term, and $Y$ is the predicted output.

During the MGGP run, in addition to the standard GP subtree crossover (a low-level crossover), genes can be acquired or deleted using a tree crossover operator called high level crossover. In the low-level crossover, a gene is chosen at random from each parent individual. Then, the standard sub-tree crossover is applied and the created trees will replace the parent trees in the otherwise unaltered individual in the next generation. The high-level crossover allows the exchange of one or more genes with another selected individual subject to the Gmax constraint. If an exchange of genes results in any individual containing more genes than Gmax, the genes will be randomly selected and deleted until the individual contains Gmax genes [21].

With respect to genetic operators, mutation in MGGP is similar to that in GP. As for crossover, the level at which the operation is performed must be specified: it is possible to apply crossover at high and low levels.

In general, the evolutionary process in MGGP differs from that in GP due to the addition of two parameters [21]: maximum number of trees per individual and high level crossover rate. A high value is normally used for the first parameter to assure a smooth evolutionary process. On the other hand, the high level crossover rate, similarly to other genetic operator rates, needs to be adjusted.

Typically, standard GP algorithms (including MGGP) will optimize only one objective in the model development process: maximizing the goodness-of-fit to the training data. However the developed models, by using a single objective in the optimization process, can become overly complex.

To overcome this drawback, a new algorithm called multi-objective genetic programming (MOGP) was introduced in [18]. The MOGP effectively combines the model structure selection ability of a standard genetic programming with the parameter estimation power of classical regression via MOGP, and it simultaneously optimizes competing two objectives the complexity and goodness-of-fit in a system (i.e. maximizing the goodness-of-fit and minimizing the model complexity) through a non-dominated sorting algorithm.

## III. The proposed method

In this paper, we propose a new metric of full reference image quality assessment based of multi-gene MOGP, where we used the open source GPTIPS2 toolbox for free genetic programming and multi-gene symbolic regression. GPTIPS and documentation are available for download[1].

For training data, a set of existing input values, we use 16 IQA measures, and corresponding output values are the subjective MOS (Mean Opinion Score). The 16 following techniques were used: VSI [22], FSIM [23], FSIMc[23], GSM [24], IFC [25], IW-SSIM [26], MAD [6], MSSIM [28], NQM [29], PSNR [30], RFSIM[31], SR-SIM [32], VIF [34], IFS [35], and SFF [36], SSIM [37].

In experiments, the following four image benchmarks were used: LIVE [37] (29 reference images, 779 distorted images),

---
[1] http://sites.google.com/site/gptips4matlab/.

CSIQ [6] (30 reference images, 886 distorted images), TID2008 [39] (25 reference images, 1700 distorted images), and TID2013 [40] (25 reference images, 3000 distorted images). Each one contains the subjective human evaluations in the form of mean opinion scores (MOS) or differential MOS (DMOS).

Parameters of the MOGP were determined experimentally observing convergence of the objectives functions over the generations. Two objectives are used here:

1) Maximise the goodness-of-fit, where we use the coefficient of determination ($R^2$) measure [41]:

$$R^2 = (obj - sub)^2 / \sum_1^m ((obj - mean(obj))^2) \quad (2)$$

Where: *Obj*: vector of objective scores,
*Sub*: MOS or differential MOS (DMOS),
*m*: the total number of images.

2) Minimize the complexity of the model, where "Expressional Complexity" is defined as the sum of nodes of all sub-trees within a tree as defined in [38]; this sort of metric has the advantage of favouring fewer layers as well as providing more resolution at the low end of the complexity axis of the Pareto front so that more simple solutions may be included in the Pareto front.

To achieve these objectives, a parametric study was designed to test the effect of the modification of each parameter.
TABLE I shows the parameter settings used for the MOGP implementation in this study.

TABLE I. PARAMETERS SETTING

| Parameter | Setting |
|---|---|
| Population size | 100 |
| Number of generations | 100 |
| Function set | +, −. |
| Maximum number of genes allowed in individual (Gmax) | 3 |
| Maximum tree depth (Dmax) | 5 |
| Tournament size | 2 |
| elite_fraction | 0.05 |
| Crossover event | 0,85 |
| Mutation events | 0,3 |

MOGP starts with an initial population of randomly generated computer programs composed of functions and terminals appropriate to the problem domain [19]. The functions used here are ''+'' and ''-'' for linear combination, and terminals are 16 full reference IQA measures.

Then, from generation to generation, after applying selection, crossover and mutation operators, better computer programs (represented by tree structures) are emerging, with Maximum tree depth is 5 and 3 genes allowed in each individual. The MOGP was run for 100 generations, with a population of 100 individuals, elite count equal to 0.05 of the population size, and 0.85 crossover fractions. Scattered crossover, Gaussian mutation and stochastic uniform selection rules were used.

Parameters of the MOGP were determined experimentally observing convergence of the objective functions (accuracy and complexity) over the generations.
All presented calculations were performed using Matlab software (version R2016a) with GPTIPS2 Toolbox.
After 100 runs on each benchmark, the proposed MFMOGP were obtained:

**MFMOGP1(LIVE)** = 207.4 *GSM* - 46.77 *FSIM* + 11.27 *IFC* + 24.24 *IWSSIM* + 68.18 *MAD* + 11.27 *NQM* - 46.77 *VIF* - 35.51 *IFS* - 114.9 (3)

**MFMOGP2(CSIQ)** = 2.012 *GSM* - 1.112 *VSI* + 0.4993 *MAD* + 0.6708 *MSSIM* - 0.1912 *RFSIM* - 0.4408 *SRSIM* - 0.2299 *VIF* - 0.4408 *IFS* + 0.2496 *SFF* - 0.5207 (4)

**MFMOGP3(TID2008)** = 10.31 *VSI* + 0.6498 *IWSSIM* - 1.958 *MAD* - 0.6498 *NQM* + 1.958 *PSNR* + 0.6589 *RFSIM* + 3.917 *SRSIM* - 3.848 *SSIM* + 1.958 *VIF* + 3.848 *IFS* - 11.37 (5)

**MFMOGP4(TID2013)** = 15.28 *VSI* - 14.09 *FSIM* + 14.09 *FSIMC* + 3.754 *GSM* - 2.565 *MAD* - 3.754 *MSSIM* - 1.189 *PSNR* + 1.189 *VIF* + 3.754 *IFS* - 13.96 (6)

The Fig 2 displays the population of evolved models in terms of their complexity as well as their fitness (1-$R^2$). Where the blue circles show the results of all developed models, the green circles comprise the pareto-optimal models in the population. And the selected models are encircled in red.

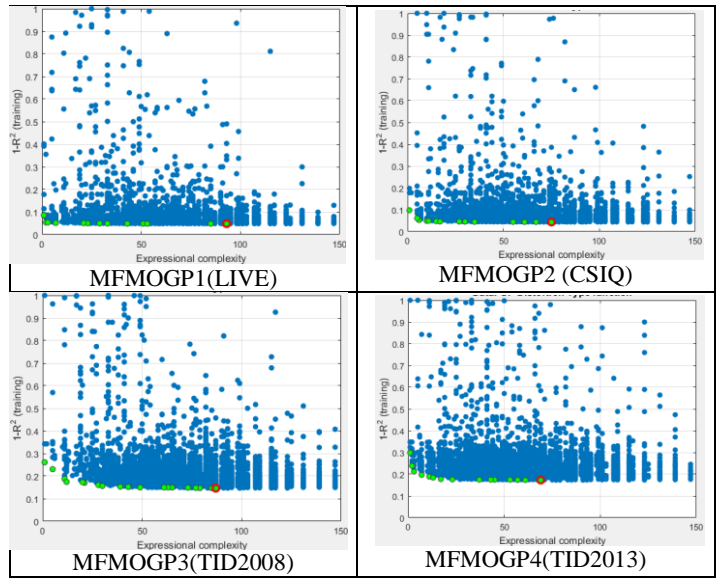

**Fig. 2** Pareto optimal fronts (model fitness vs. complexity) of MFMOGPs

## IV. Experimental evaluation

To check the efficiency of the proposed approach, we use a MOS scatter plot versus the results obtained; where the diagram should have a compact form without outliers and with a tendency to a monotonic behaviour.

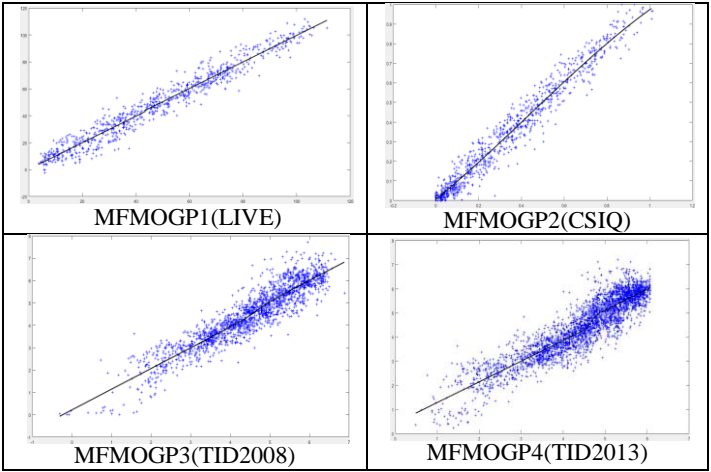

**Fig. 3** Correlations of subjective and objective assessment obtained (Each data point represent one test image).

Fig. 3 shows the scatter plots of the score given by the result of our new method versus the subjective MOS. additionally; a fit with a logistic function as suggested in [30] is shown for easier comparison.

We note that the result is adequate to human perception; where the percentage of outliers is decreased and a tendency to a monotonic behaviour is increased.

For evaluation and development of four developed MFMOGPs, the four indices of prediction accuracy, monotonicity, and consistency are used [27],[30]: Root Mean Square Error (RMSE), Pearson linear Correlation Coefficient (PCC), Spearman Rank order Correlation Coefficient (SRCC) and Kendall Rank order Correlation Coefficient (KRCC).

Table II presents evaluation results, for the best ten models (Among 16 metrics used) and MFMOGP. The top two models for each criterion are shown in boldface.

TABLE II. COMPARISON OF PERFORMANCE OF THE PROPOSED APPROACH WITH THE BEST TEN MODELS OF IQA

| TID13 | VSI | IFS | SFF | FSIMc | SRSIM | FSIM | GSM | MAD | MSSIM | VIF | MFMOGP1 | MFMOGP2 | MFMOGP3 | MFMOGP4 |
|---|---|---|---|---|---|---|---|---|---|---|---|---|---|---|
| SRCC | **0.8965** | 0.8697 | 0.8513 | 0.8510 | 0.7999 | 0.8015 | 0.7946 | 0.7807 | 0.7859 | 0.6769 | 0.8139 | 0.8347 | 0.8677 | **0.9019** |
| KRCC | **0.7183** | 0.6785 | 0.6581 | 0.6665 | 0.6314 | 0.6289 | 0.6255 | 0.6035 | 0.6047 | 0.5147 | 0.6359 | 0.6554 | 0.6878 | **0.7252** |
| PCC | **0.9000** | 0.8791 | 0.8706 | 0.8769 | 0.8590 | 0.8589 | 0.8464 | 0.8267 | 0.8329 | 0.7735 | 0.8517 | 0.8773 | 0.8832 | **0.9140** |
| RMSE | **0.5404** | 0.5909 | 0.6099 | 0.5959 | 0.6347 | 0.6349 | 0.6603 | 0.6975 | 0.6861 | 0.7856 | 0.6497 | 0.5950 | 0.5814 | **0.5029** |

| TID08 | VSI | SRSIM | IFS | FSIMc | FSIM | SFF | GSM | MAD | MSSIM | VIF | MFMOGP1 | MFMOGP2 | MFMOGP3 | MFMOGP4 |
|---|---|---|---|---|---|---|---|---|---|---|---|---|---|---|
| SRCC | 0.8979 | 0.8913 | 0.8903 | 0.8840 | 0.8805 | 0.8767 | 0.8504 | 0.8340 | 0.8542 | 0.7491 | 0.8974 | **0.9119** | **0.9230** | 0.9065 |
| KRCC | 0.7123 | 0.7149 | 0.7009 | 0.6991 | 0.6946 | 0.6882 | 0.6596 | 0.6445 | 0.6568 | 0.5860 | 0.7169 | **0.7407** | **0.7576** | 0.7271 |
| PCC | 0.8762 | 0.8866 | 0.8810 | 0.8762 | 0.8738 | 0.8817 | 0.8422 | 0.8306 | 0.8451 | 0.8084 | 0.9035 | **0.9117** | **0.9246** | 0.9018 |
| RMSE | 0.6466 | 0.6206 | 0.6349 | 0.6468 | 0.6525 | 0.6333 | 0.7235 | 0.7473 | 0.7173 | 0.7899 | 0.5752 | **0.5514** | **0.5111** | 0.5798 |

| CSIQ | VSI | SRSIM | IFS | SFF | MAD | FSIM | FSIMc | GSM | MSSIM | VIF | MFMOGP1 | MFMOGP2 | MFMOGP3 | MFMOGP4 |
|---|---|---|---|---|---|---|---|---|---|---|---|---|---|---|
| SRCC | 0.9423 | 0.9319 | 0.9582 | 0.9627 | 0.9466 | 0.9242 | 0.9310 | 0.9108 | 0.9133 | 0.9195 | **0.9652** | **0.9735** | 0.9630 | 0.9649 |
| KRCC | 0.7857 | 0.7725 | 0.8165 | 0.8288 | 0.7970 | 0.7567 | 0.7690 | 0.7374 | 0.7393 | 0.7537 | **0.8377** | **0.8585** | 0.8313 | 0.8349 |
| PCC | 0.9279 | 0.9250 | 0.9576 | 0.9643 | 0.9500 | 0.9120 | 0.9192 | 0.8964 | 0.8991 | 0.9277 | **0.9721** | **0.9782** | 0.9673 | 0.9696 |
| RMSE | 0.0979 | 0.0997 | 0.0757 | 0.0695 | 0.0820 | 0.1077 | 0.1034 | 0.1164 | 0.1149 | 0.0980 | **0.0616** | **0.0545** | 0.0666 | 0.0642 |

| LIVE | FSIM | FSIMc | MAD | SRSIM | SFF | VSI | GSM | MSSIM | VIF | IFS | MFMOGP1 | MFMOGP2 | MFMOGP3 | MFMOGP4 |
|---|---|---|---|---|---|---|---|---|---|---|---|---|---|---|
| SRCC | 0.9634 | 0.9645 | 0.9669 | 0.9618 | 0.9649 | 0.9524 | 0.9561 | 0.9513 | 0.9636 | 0.9599 | **0.9751** | **0.9712** | 0.9687 | 0.9660 |
| KRCC | 0.8337 | 0.8363 | 0.8421 | 0.8299 | 0.8365 | 0.8058 | 0.8150 | 0.8045 | 0.8282 | 0.8254 | **0.8605** | **0.8490** | 0.8416 | 0.8347 |
| PCC | 0.9597 | 0.9613 | 0.9675 | 0.9553 | 0.9632 | 0.9482 | 0.9512 | 0.9489 | 0.9411 | 0.9586 | **0.9758** | **0.9701** | 0.9676 | 0.9651 |
| RMSE | 7.6781 | 7.5297 | 6.9073 | 8.0813 | 7.346 | 8.6816 | 8.4327 | 8.6188 | 9.2402 | 7.7765 | **5.9799** | **6.6341** | 6.9011 | 7.1565 |

Results shows that all MFMOGP measures are clearly a top performing models compared measures on for databases, where MFMOGP4 significantly outperformed compared measures on TID 2013 databases. MFMOGP2 yielded good results on LIVE and TID2008 databases but the best results is on CSIQ one. MFMOGP1 performed well on LIVE, it was also the second best measure on CSIQ. MFMOGP3, in turn, yielded best results on TID2008 databases.

TABLE III. COMPARISON OF THE APPROACH WITH OTHER FUSION IQA MEASURES BASED ON SRCC VALUES

| IQA measure | TID2013 | TID2008 | CSIQ | LIVE |
|---|---|---|---|---|
| **MAD [6]** | 0,7807 | 0,8340 | 0,9466 | 0,9669 |
| **CQM [8]** | - | 0,8720 | - | - |
| **Lahouhou et al [11]** | - | - | - | 0,9500 |
| **ADM[46]** | - | 0,8617 | 0,9333 | 0,9460 |
| **BMMF [42]** | 0,8340 | 0,9471 | - | - |
| **BME [12]** | - | 0,8882 | 0,9573 | 0,9711 |
| **RMSSIM[13]** | - | 0,8569 | 0,9453 | 0,9633 |
| **IGN[43]** | - | 0,8902 | 0,9401 | 0,9580 |
| **EHIS[9]** | - | 0,9098 | 0,9498 | 0,9622 |
| **MMF[33]** | - | **0,9487** | **0,9755** | 0,9732 |
| **GLD-PFT[7]** | - | 0,8849 | 0,9549 | 0,9631 |
| **Barri rt al[44]** | - | 0,8100 | 0,9630 | 0,9570 |
| **DOG-SSIM[45]** | 0,8942 | 0,9259 | 0,9204 | 0,9423 |
| **ESIM[16]** | 0,8804 | 0,9026 | 0,9620 | 0,9420 |
| **LCSIM[17]** | **0,9044** | 0,9178 | 0,9733 | 0,9749 |
| **MFMOGP1** | 0,8139 | 0,8974 | 0,9652 | **0,9751** |
| **MFMOGP2** | 0,8677 | 0,9119 | **0,9735** | 0,9712 |
| **MFMOGP3** | 0,8677 | **0,9230** | 0,9630 | 0,9687 |
| **MFMOGP4** | **0,9019** | 0,9065 | 0,9649 | 0,9660 |

It would be agreeable to compare the proposed measure with other related fusion IQA measures. Table III contains such comparative evaluation based on SRCC values. Since values of PCC, KRCC and RMSE are often not available, SRCC was used as a basis for comparison. Two best results for a given benchmark dataset are written in boldface, some results were not reported in referred works; therefore, they are denoted by "-".

Evaluation results show that for LIVE database, the tow measures MFMOGP1 and MFMOGP2 outperformed other approaches. After MMF, MFMOGP3 measure is a best technique on TID2008 database, and MFMOGP2 is a best one on CSIQ dataset. On TID2013 database, LCSIM provided good results and MFMOGP4 is the second best measures.

# Conclusion

In this paper, a fusion full-reference IQA was introduced, in which the quality of a given image is evaluated in a way that is consistent with human evaluation, by a well-performing hybrid measure which consists of a small number of IQA measures.

The obtained fusion was able to find the appropriate IQA measures from 16 full-reference IQA measures, used as predictors in the multiple linear regression, by optimizing simultaneously two competing objectives of model 'goodness of fit' to data and model complexity.

For evaluation, four largest image benchmarks (TID2013, TID2008, CSIQ, and LIVE) and four performance indices (SRCC, PCC, KROCC, RMSE) were used.

Four IQA measures, called Multi-measures Fusion based on Multi-Objective Genetic programming (MFMOGPs), were obtained, in experiments. The presented results confirm good performance of obtained MFMOGPs in comparison with state-of-the-art IQA techniques, including other recently published fusion approaches. In future works, we will consider incorporating other IQA measures for fusion.

**REFERENCES**

[1] Chandler D. M. "Seven challenges in image quality assessment: past, present, and future research", ISRN Signal Processing, 2013.

[2] Liu T. J., Lin Y. C., Lin W. & Kuo, C. C. J., "Visual quality assessment: recent developments, coding applications and future trends", APSIPA Transactions on Signal and Information Processing, 2013, 2, e4.

[3] S. Ryu, D.H. Kim, K. Sohn, "Stereoscopic image quality metric based on binocular perception model," IEEE International Conference on Image Processing, pp. 609-612, 2012.

[4] C.T.E.R. Hewage and M.G. Martini, "Reduced-reference quality metric for 3d depth map transmission," in 3DTV-CON, pp. 1–4, 2010.

[5] Z. Wang, H. R. Sheikh & A. C. Bovik, "No-reference perceptual quality assessment of JPEG compressed images". In Proceedings of IEEE International Conference on Image Processing, volume 1, pages 477–480. (2002).

[6] Larson EC, Chandler DM (2010), "Most apparent distortion: full-reference image quality assessment and the role of strategy". J Electron Imaging 19(1):011006.

[7] Saha A, Wu QMJ (2014) Full-reference image quality assessment by combining global and local distortion measures. CoRR abs/1412.5488

[8] Okarma K (2010) Combined full-reference image quality metric linearly correlated with subjective assessment. In: Artificial intelligence and soft computing. Springer, pp 539–546.

[9] Okarma K (2013) Extended Hybrid Image Similarity - combined full-reference image quality metric linearly correlated with subjective scores. Elektronika ir Elektrotechnika 19(10):129–132

[10] Liu M, Yang X (2009) A new image quality approach based on decision fusion. In: Ma J, Yin Y, Yu J, Zhou S (eds) Proc. Int. conf. on fuzzy systems and knowledge discovery (FSKD). IEEE, pp 10–14

[11] Lahouhou A, Viennet E, Beghdadi A (2010) Selecting low-level features for image quality assessment by statistical methods. CIT 18(2)

[12] Peng P, Li ZN (2012) A mixture of experts approach to multi-strategy image quality assessment. In: Campilho A, KamelM(eds) Image analysis and recognition, lecture notes in computer science, vol 7324. Springer, Berlin-Heidelberg, pp 123–130,

[13] Peng P, Li ZN (2012) Regularization of the structural similarity index based on preservation of edge direction. In: 2012 IEEE International conference on systems, man, and cybernetics (SMC), pp 2127–2132.

[14] Lukin VV, Ponomarenko NN, Ieremeiev OI, Egiazarian KO, Astola J (2015) Combining full-reference image visual quality metrics by neural network. In: Human vision and electronic imaging XX, Proc. SPIE, vol 9394, p 93940K.

[15] Yuan Y, Guo Q, Lu X (2015) Image quality assessment: a sparse learning way. Neurocomputing 159:227–241.

[16] Oszust M (2016) Decision fusion for image quality assessment using an optimization approach. IEEE Signal Proc Let 23(1):65–69.

[17] Oszust M (2016) Full-Reference Image Quality Assessment with Linear Combination of Genetically Selected Quality Measures. PLoS ONE 11(6): e0158333.

[18] A.H. Gandomi, et al., Genetic programming for experimental big data mining: A case study on concrete creep formulation, Automation in Construction(2016),

[19] Koza J.R., ''Genetic programming: on the programming of computers by means of natural selection'', The MIT Press, USA, (1992).

[20] Poli R., Langdon W.B. & McPhee N.F., A field guide to genetic programming, Published via http://lulu.com and freely available at http://www.gp-field-guide.org.uk, 2008.

[21] Charles Hii, et al., ''Evolving Toxicity Models using Multigene Symbolic Regression and Multiple Objectives''. International Journal of Machine Learning and Computing, Vol.1, No. 1, April 2011

[22] Zhang L, Shen Y, Li H. VSI: ''A visual saliency-induced index for perceptual image quality assessment''. IEEE T Image Process. 2014 Oct; 23(10):4270–4281.

[23] Zhang L, Zhang L, Mou X, Zhang D. FSIM: ''A feature similarity index for image quality assessment''. IEEE T Image Process. 2011 Aug; 20(8):2378–2386.

[24] Liu A, Lin W, Narwaria M. ''Image quality assessment based on gradient similarity''. IEEE T Image Process. 2012 Apr; 21(4):1500–1512.

[25] Sheikh HR, Bovik AC, de Veciana G. ''An information fidelity criterion for image quality assessment using natural scene statistics''. IEEE T Image Process. 2005 Dec; 14(12):2117–2128.

[26] Wang Z, Li Q. ''Information content weighting for perceptual image quality assessment''. IEEE T Image Process. 2011 May; 20(5):1185–1198.

[27] VQEG: Final Report from the video quality experts group on the validation of objective models of video quality assessment, FR-TV Phase II, http://www.vqeg.org/

[28] Wang Z, Simoncelli EP, Bovik AC. ''Multi-scale structural similarity for image quality assessment''. In:Proc. IEEE Int. Conf. on Signals, Systems, and Computers, (ASILOMAR); 2003. p. 1398–1402.

[29] Damera-Venkata N, Kite TD, Geisler WS, Evans BL, Bovik AC. ''Image quality assessment based on a degradation model''. IEEE T Image Process. 2000 Apr; 9(4):636–650.

[30] Hamid Rahim Sheikh, Muhammad Farooq Sabir, and Alan Conrad Bovik.: ''A Statistical Evaluation of Recent Full Reference Image Quality Assessment Algorithms''. IEEE transactions on image processing, vol. 15, no.11, pp.3440 - 3451 (2006)

[31] Zhang L, Zhang L, Mou X. RFSIM: ''A feature based image quality assessment metric using Riesz transforms''. In: Proc. IEEE Int. Conf. on Image Processing (ICIP). IEEE; 2010.

[32] Zhang L, Li H. ''SR-SIM: A fast and high performance IQA index based on spectral residual''. In: Proc. IEEE Int. Conf. on Image Processing (ICIP). IEEE; 2012.

[33] Liu TJ, Lin W, Kuo CC (2013) Image quality assessment using multi-method fusion. IEEE T Image Process 22(5):1793–1807.

[34] Sheikh HR, Bovik AC. ''Image information and visual quality''. IEEE T Image Process. 2006 Feb; 15 (2):430–444.

[35] Chang HW, Zhang QW, Wu QQ, Gan Y. ''Perceptual image quality assessment by independent feature detector''. Neurocomputing. 2015; 151, part 3:1142–1152. doi: 10.1016/j.neucom.2014.04.081

[36] Chang HW, Yang H, Gan Y, Wang MH. ''Sparse feature fidelity for perceptual image quality assessment''. IEEE T Image Process. 2013 Oct; 22(10):4007–4018.

[37] Z. Wang, A. C. Bovik, H. R. Sheikh, and E. P. Simoncelli, "Image quality assessment: From error visibility to structural similarity," IEEE Trans. Image Process., vol. 13, no. 4, pp. 600–612, Apr. 2004.

[38] Guido F. Smits and Mark Kotanchek,"Pareto-front exploitation in symbolic regression", pp. 283 - 299, Genetic Programming Theory and Practice II, 2004.

[39] N. Ponomarenko, V. Lukin, A. Zelensky, K. Egiazarian, M. Carli, and F. Battisti, "TID2008 - a database for evaluation of full-reference visual quality assessment metrics," Adv. Mode. Radioelectron., vol. 10, pp. 30–45, 2009.

[40] N. Ponomarenko, L. Jin, O. Ieremeiev, V. Lukin, K. Egiazarian, J.Astola, B. Vozel, K. Chehdi, M. Carli, F. Battisti, and C.-C. J. Kuo, "Image database TID2013: Peculiarities results and perspectives," Signal Process.-Image, vol. 30, pp. 57–77, Jan. 2015.

[41] Gerald J. Hahn. ''The coefficient of determination exposed!''. Reprinted from CHEMICAL TECHNOLOGY. VOl3. No. 10. October 1973. Page 609

[42] Jin L, Egiazarian K, Kuo CCJ. Perceptual image quality assessment using block-based multi-metric fusion (BMMF). In: Proc. IEEE Int. Conf. on Acoustics, Speech and Signal Processing (ICASSP); 2012. p. 1145–1148.

[43] Wu J, Lin W, Shi G, Liu A. Perceptual quality metric with internal generative mechanism. IEEE T Image Process. 2013 Jan; 22(1):43–54.

[44] Barri A, Dooms A, Jansen B, Schelkens P. A locally adaptive system for the fusion of objective quality measures. IEEE T Image Process. 2014 Jun; 23(6):2446–2458.

[45] Pei SC, Chen LH. Image quality assessment using human visual DOG model fused with random forest. IEEE T Image Process. 2015 Nov; 24(11):3282–3292.

[46] Li S, Zhang F, Ma L, Ngan KN. Image quality assessment by separately evaluating detail losses and additive impairments. IEEE T Multimedia. 2011 Oct; 13(5):935–949.